\documentstyle[aps,prl,preprint,floats,epsfig]{revtex}

\begin{document}

\preprint{\tighten\vbox{\hbox{\hfil CLNS 01/1740}
                        \hbox{\hfil CLEO 01-12}
}}

\title{First Measurement of $\Gamma(D^{\ast +})$}  

\author{CLEO Collaboration}
\date{7 August 2001}

\maketitle
\tighten

\begin{abstract} 
We present the first measurement of the $D^{\ast +}$ width using
9/fb of $e^+e^-$ data collected near the $\Upsilon(4\rm{S})$
resonance by the CLEO II.V detector.
Our method uses advanced tracking techniques
and a reconstruction method that takes advantage of the small
vertical size of the CESR beam spot to measure the energy release
distribution from the $D^{\ast +} \to D^0 \pi^+$ decay.
We find $\Gamma(D^{\ast +}) = 96 \pm 4\ ({\rm Statistical})
\pm 22\ ({\rm Systematic})$ keV.  We also measure the energy
release in the decay and compute $\Delta m \equiv 
m_{D^{\ast +}} - m_{D^0} = 145.412 \pm 0.002 ({\rm Statistical})
\pm 0.012\ ({\rm Systematic})$ MeV/c$^2$.
\end{abstract}

\newpage

{
\renewcommand{\thefootnote}{\fnsymbol{footnote}}

\begin{center}
S.~Ahmed,$^{1}$ M.~S.~Alam,$^{1}$ S.~B.~Athar,$^{1}$
L.~Jian,$^{1}$ L.~Ling,$^{1}$ M.~Saleem,$^{1}$ S.~Timm,$^{1}$
F.~Wappler,$^{1}$
A.~Anastassov,$^{2}$ E.~Eckhart,$^{2}$
% ...
\end{center}
 
\small
\begin{center}
$^{1}${State University of New York at Albany, Albany, New York 12222}\\
$^{2}${Ohio State University, Columbus, Ohio 43210}\\
% ...
$^{14}${California Institute of Technology, Pasadena, California 91125}\\
% ...
\end{center}

\setcounter{footnote}{0}
}
\newpage

A measurement of $\Gamma(D^{\ast +})$ opens an important window
on the non-perturbative strong physics involving heavy quarks.
The basic framework of the theory is well understood, however, there is
still much speculation -
predictions for the width range from $15\,\rm{keV}$ to 
$150\,\rm{keV}$ \cite{pred}.
We know the $D^{\ast +}$ width is dominated by strong decays.
The level splitting in the $B$ sector
is not large enough to allow real strong transitions.
Therefore, a measurement of the width of the $D^{\ast +}$
gives unique information about the
strong coupling constant in heavy-light meson systems.
This width only depends on $g$,
a universal strong coupling between heavy
vector and pseudoscaler mesons to the pion, since the
small contribution of the electromagnetic decay can be neglected,
yeilding
\begin{equation}
\Gamma(D^{\ast +}) = \frac{2g^2}{12\pi f_\pi^2}p^3_{\pi^+} +
                     \frac{ g^2}{12\pi f_\pi^2}p^3_{\pi^0},
\end{equation}
where $f_\pi$ is the pion decay constant and the momenta are
for the indicated particle in $D^{\ast +}$ decay
in the $D^{\ast +}$ rest frame \cite{wise}.

Prior to this measurement, the $D^{\ast +}$ width was limited
to be less than $131\,\rm{keV}$ at the $90\%$ confidence level 
by the ACCMOR collaboration \cite{ACCMOR}.
This letter describes
a measurement of the $D^{\ast +}$ width with the CLEO II.V detector~\cite{PRD}.
The signal is
reconstructed through a single, well-measured sequence,
$D^{\ast +} \rightarrow \pi^+_{\rm slow} D^0$,
$D^0 \rightarrow K^-\pi^+$.  Consideration of
charge conjugated modes are implied throughout this letter.

	The CLEO detector has been described in detail elsewhere.  All of
the data used in this analysis are taken with the detector in its
II.V configuration \cite{CLEO}.
The data were taken in symmetric $e^+e^-$ collisions at a center of
mass energy around 10 GeV with an integrated luminosity of 9.0/fb provided by
the Cornell Electron-positron Storage Ring (CESR).  The nominal sample
follows the selection of
$D^{\ast +} \to \pi_{\rm slow}^+ D^0 \to K^-\pi^+\pi_{\rm slow}^+$ candidates
used in our $D^0-\bar{D^0}$ mixing analysis\cite{Dmix}.

\begin{figure}
     \centerline{\epsfig{file=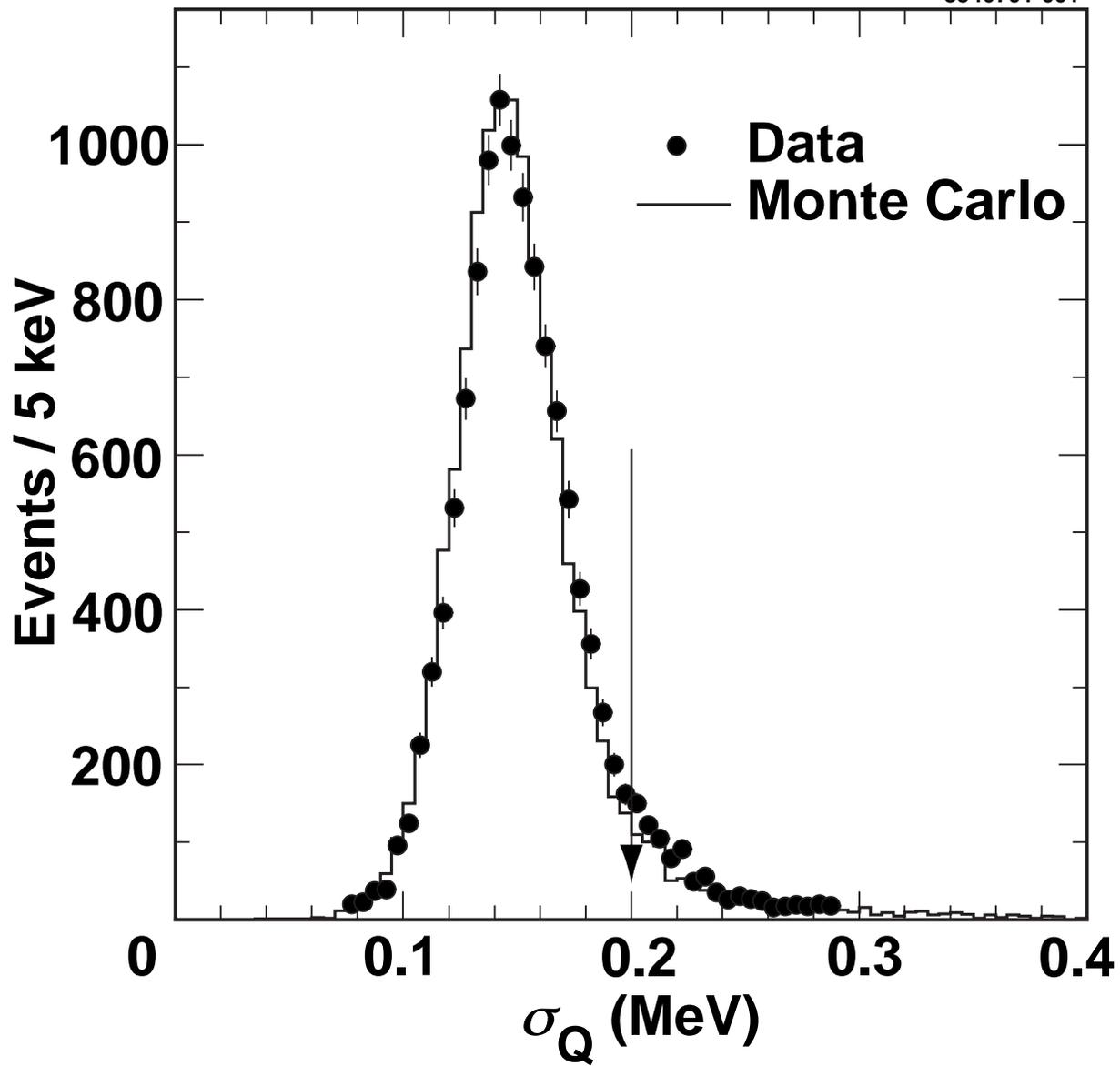,width=\linewidth}}
     \caption{\label{fig:errorcompare} Distribution of $\sigma_Q$,
              the uncertainty on $Q$ as determined from propagating
              track fitting errors.  The arrow indicates a selection
              discussed in the text.}
\end{figure}

	Our reconstruction method takes advantage of the small CESR beam
spot and the kinematics and topology of the
$D^{\ast +} \to \pi^+_{\rm slow} D^0 \to \pi^+_{\rm slow} K^- \pi^+$
decay chain. The $K^-$ and $\pi^+$ are required to form a common vertex.
The resultant $D^0$ candidate momentum vector is then projected back to the
CESR luminous region to determine the $D^0$ production point.  
The CESR luminous region has a Gaussian width $\sim 10\ \mu$m vertically and
$\sim 300\ \mu$m horizontally.
This procedure determines an accurate
$D^0$ production point for $D^0$'s moving out of the horizontal plane;
$D^0$'s moving within 0.3 radians of the horizontal plane are not
considered.  Then the $\pi_{\rm slow}^+$ track
is refit constraining its trajectory to intersect the $D^0$
production point.  This improves the resolution on the energy release,
$Q = M(K^-\pi^+\pi_{\rm slow}^+) - M(K^-\pi^+) - m_{\pi^+}$, by more
than 30\% over simply forming the appropriate invariant masses of the tracks.
The improvement to resolution is essential to our measurement of the width
of the $D^{\ast +}$.  The distribution of our
resolution, $\sigma_Q$, is shown in Figure~\ref{fig:errorcompare}
and is typically 150 keV.
The good agreement between 
Monte Carlo and data demonstrates that the kinematics and sources of 
uncertainties on the tracks, such as the number of hits used and the
effects of multiple scattering in detector material, are well modeled.

The challenge of measuring the width of the $D^{\ast +}$ is understanding the tracking
system response function since the experimental resolution exceeds the width
we are trying to measure. We depend
on exhaustive comparisons between a GEANT \cite{GEANT} based detector
simulation and our data.  We addressed the problem by selecting samples of
candidate $D^{\ast +}$ decays using three strategies.

	First we produced the largest sample from data and simulation by 
imposing only
basic tracking consistency requirements.  We call this the
{\em nominal} sample.
Second we refine the nominal sample selecting candidates
with the best measured tracks by making very tight cuts
on tracking parameters.
We call this the {\em tracking selected} sample.
A third alternative is to select our data on specific kinematic
properties of the $D^{\ast +}$ decay that minimize the dependence
of the width of the $D^{\ast +}$ on detector mismeasurements.
We call this the {\em kinematic selected} sample.
In all three samples the width is extracted with an unbinned maximum
likelihood fit to the energy release distribution and compared with
the simulation's generated value to determine a bias which is
then applied to the data.
These three different approaches yield consistent values for
the width of the $D^{\ast +}$ giving us confidence that our simulation
accurately models our data.

	To further improve the quality of reconstruction in our sample, we
apply some selections at the kinematic boundaries of
$\pi_{\rm slow}^+ $ momentum and
the opening angle $\theta$ between
the $\pi_{\rm slow}^+$ and the $D^0$ candidate as a function of the
$D^{\ast +}$ candidate momentum distributions
to remove a small amount of misreconstructed
signal and background.
We also require $\sigma_Q < 200$ keV which removes the long tail in
the error distribution.

	Table~\ref{tab:data} summarizes the statistics in our three samples.
The tracking and kinematic samples are subsets of the nominal sample.
The two subsets contain 94 common candidates.

	We assume that the intrinsic width of the $D^0$ is negligible,
$\Gamma(D^0) \ll \Gamma(D^{\ast +})$, implying that the width of $Q$
is simply a convolution of the shape given by the $D^{\ast +}$ width and
the tracking system response function.  Thus we
consider the pairs of $Q$ and $\sigma_Q$ for
$D^{\ast +} \to \pi_{\rm slow}^+ D^0 \to K^-\pi^+\pi_{\rm slow}^+$
where $\sigma_Q$ is given for
each candidate by propagating the tracking errors in the kinematic
fit of the charged tracks.  We perform an unbinned maximum likelihood fit
to the $Q$ distribution.

	The underlying signal shape of the $Q$ distribution is assumed to be
given by a P-wave Breit-Wigner
with central value $Q_0$.
We considered a relativistic and non-relativistic
Breit-Wigner as a model of the
underlying signal shape, and found negligible changes in the fit
parameters between the two.
The width of the signal Breit-Wigner
depends on $Q$ and is given by
\begin{equation}
\Gamma(Q) = \Gamma_0  \left(\frac{P}{P_0}\right)^3 \left(\frac{M_0}{M}\right)^2 ,
\label{eq:BW}
\end{equation}
where $\Gamma_0 \equiv \Gamma(D^{\ast +})$,
$P$ and $M$ are the candidate
$\pi_{\rm slow}^+$ or $D^0$ momentum in the $D^{\ast +}$ rest frame
and $K\pi\pi_{\rm slow}$ mass, and $P_0$ and $M_0$ are
the values computed using $Q_0$.
The effect of the mass term is negligible at our energy. 
The partial width and the total width differ negligibly in  
their dependence on $Q$ for $Q>1~MeV$.

	For each candidate the signal shape is convolved with
a resolution Gaussian with width $\sigma_Q$, determined by the tracking
errors, as a model of our finite resolution shown in
Figure~\ref{fig:errorcompare}.
The fit also includes a background contribution
with a fixed shape derived from our simulation, and
modeled with a third order polynomial.
We allow a small fraction of the signal, $f_{mis}$, to be parametrized
by a single Gaussian resolution function of width $\sigma_{mis}$.
This shape is included in the fit to model the tracking
mishaps which our simulation predicts to be at the 5\% level in the
nominal sample and negligible in both the tracking
and kinematic selected samples.  In our standard fit we constrain the level
of this contribution while allowing $\sigma_{mis}$ to float.

	As a preliminary test to fitting the data we run the complete analysis
on a fully simulated sample that has about ten times the data
statistics and is generated with a range of underlying $\Gamma(D^{\ast +})$
from 0~to 130~keV.  We do this for the three samples and
compute offsets between the generated and fit values for the
width and mean energy release.
We also note that in all three simulated samples there are no trends
in the difference between measured and generated width as a function
of the generated width; the offset is consistent with zero as a function of
the generated width of the $D^{\ast +}$.
Table~\ref{tab:data} summarizes this simulation
study.  We apply these offsets to the fit value that we obtain
from the data.  For the energy release all samples show small shifts,
$-7 \pm 3$ keV for the nominal, $-12 \pm 10$ keV for the tracking,
and  $-12 \pm 5$ keV for the kinematic. 

\begin{figure}
\begin{tabular}{ccc}
\epsfxsize=55mm
\epsfbox{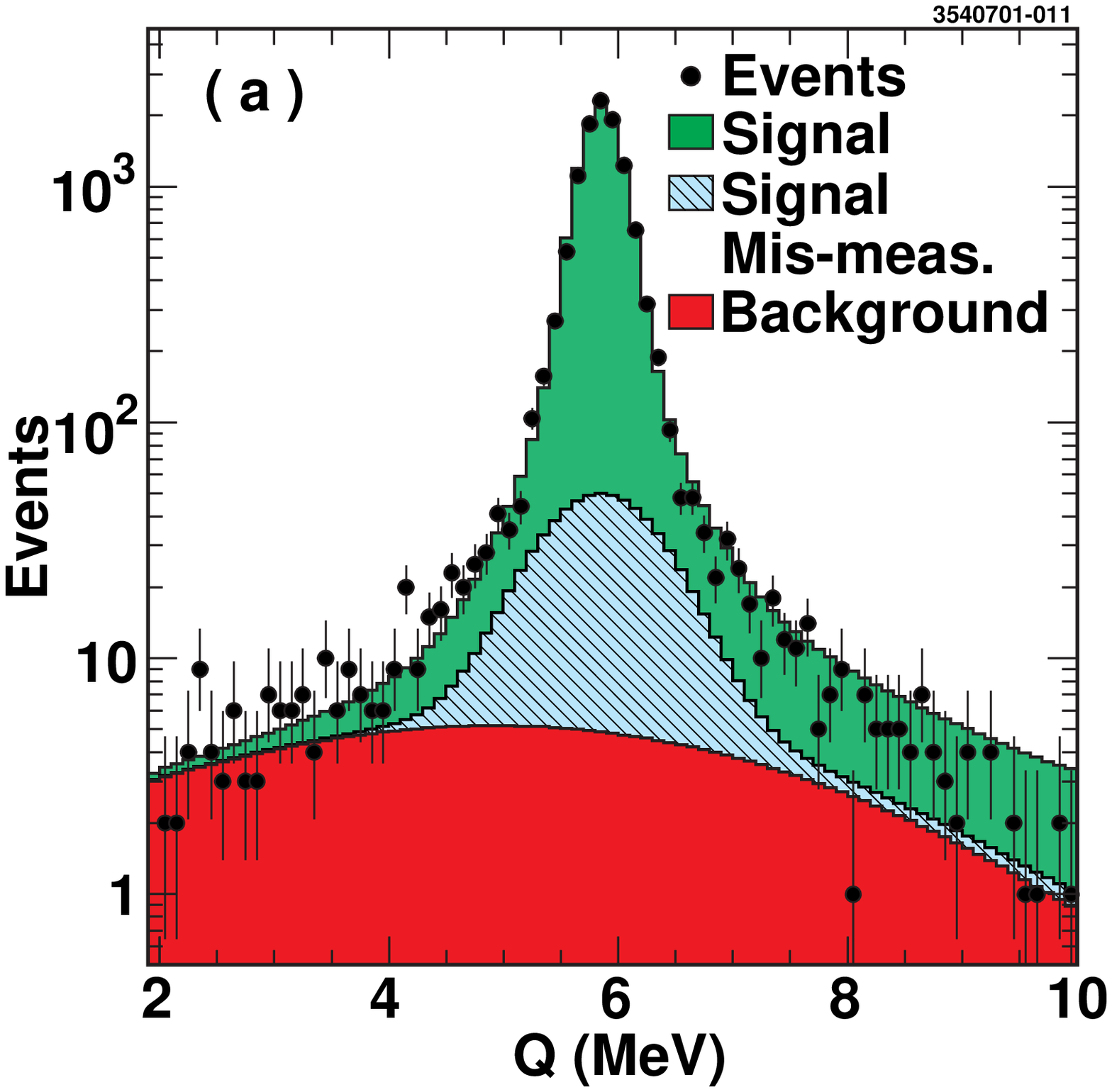} &
\epsfxsize=55mm
\epsfbox{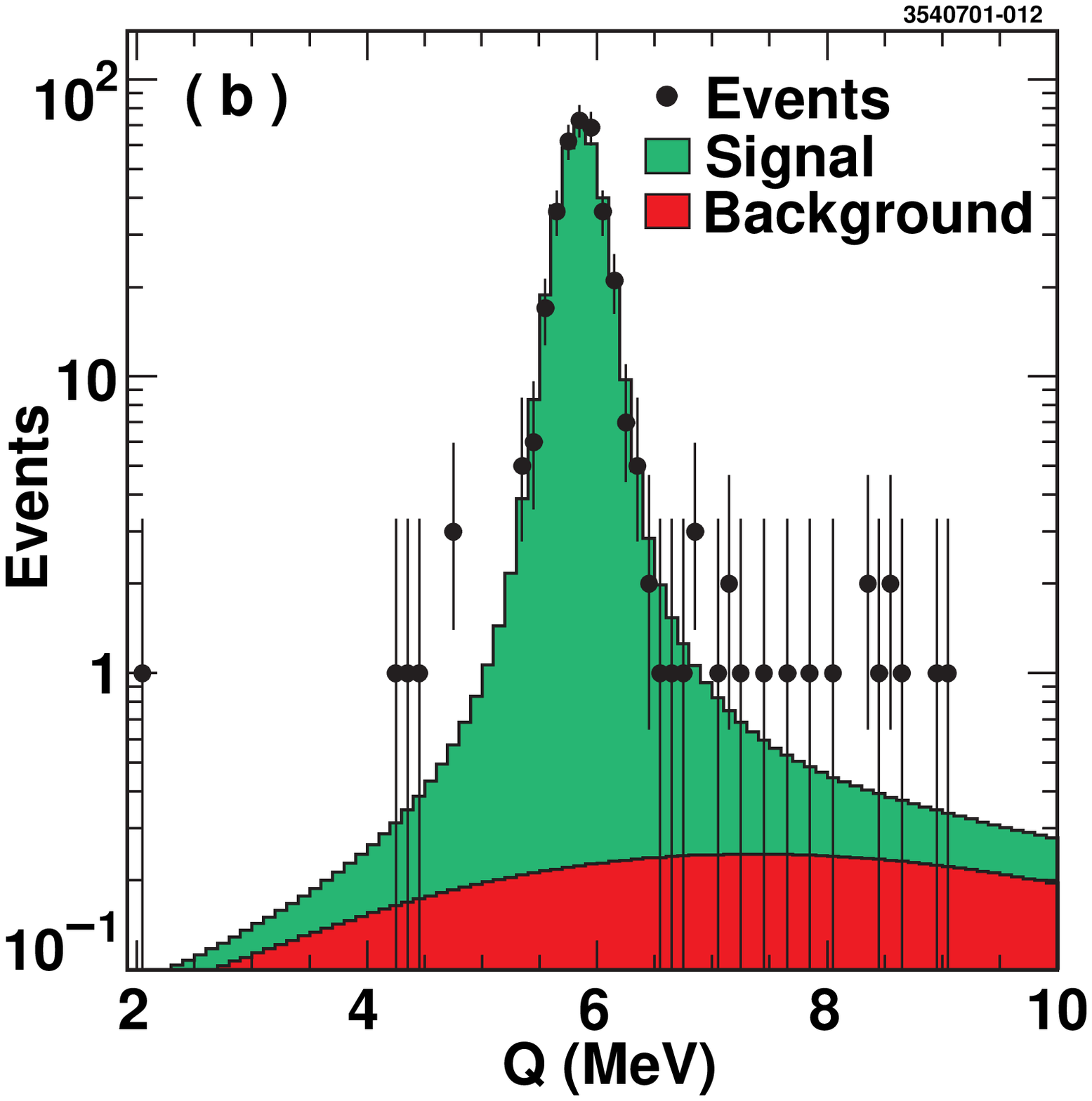} &
\epsfxsize=55mm
\epsfbox{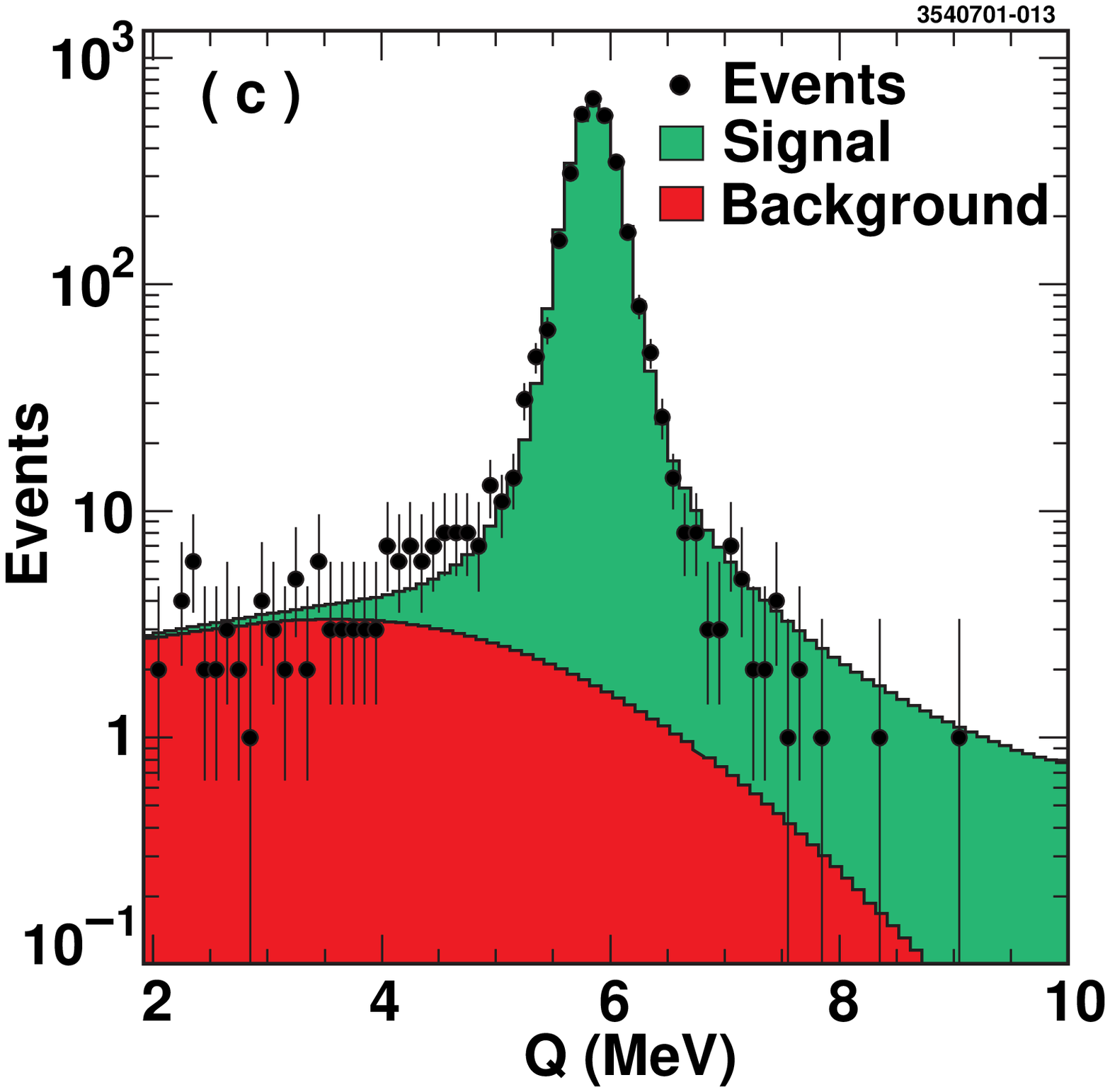}
\end{tabular}
   \caption{\label{fig:nomfit} Fits to the three data samples: a)
is nominal; b) is tracking; and c) is kinematic.  The different
contributions to the fits are shown by different colors.}
\end{figure}

	Figure~\ref{fig:nomfit}
displays the fits to the three data sample.
The results of the fits
are summarized in Table~\ref{tab:fit}.
\begin{table}
\caption{Results of the fits described in the text.  The uncertainties are
statistical.}
\begin{center}
\begin{tabular}{|c|c|c|c|} \hline
                     & \multicolumn{3}{c|}{Sample} \\ \hline
Parameter            & Nominal          & Tracking          & 
Kinematic \\ \hline
$\Gamma_0$ (keV)     & $98.9 \pm 4.0$   & $106.0 \pm 19.6$  & $108.1 
\pm  5.9$ \\
$Q_0$ (keV)          & $5853 \pm 2$     & $5854 \pm 10$     & $5850 \pm 4$ \\
$N_s$                & $11207 \pm 109$  & $353 \pm 20$      & $3151 \pm 57$ \\
$f_{mis}$ (\%)       & $5.3 \pm 0.5$    & NA                & NA \\
$\sigma_{mis}$ (keV) & $508 \pm 39 $    & NA                & NA \\
$N_b$                & $289 \pm 31 $    & $15 \pm 7$        & $133 \pm 16$ \\ \hline
\end{tabular}
\end{center}
\label{tab:fit}
\end{table}
Correlations among the floating parameters of the fit are negligible.

	The agreement is excellent among the three fits, and when the offsets
from Table~\ref{tab:data} are applied we obtain the results given in the 
last row.
\begin{table}
\caption{Summary of our data sample, simulation biases, and fit results.}
\begin{center}
\begin{tabular}{|c|c|c|c|} \hline
                     & \multicolumn{3}{c|}{Sample} \\ \hline
Parameter            & Nominal          & Tracking          & Kinematic \\ \hline
Candidates           & 11496            & 368               & 3284 \\
Background Fraction (\%)
                     & $2.51 \pm 0.27$  & $4.1 \pm 1.9$     & $4.05 \pm 0.49$ \\
$\Gamma_{\rm fit} - \Gamma_{\rm generated}$ (keV)
                     & $2.7 \pm 2.1$    & $1.7 \pm 6.4$     & $4.3 \pm 3.1$ \\
Fit $\Gamma_0$ (keV) &  $98.9 \pm 4.0$  & $106.0 \pm 19.6$ & $108.1 \pm  5.9$ \\       
$D^{\ast +}$ Width (keV) & $96.2 \pm 4.0$
                         & $104 \pm 20$
                         & $103.8 \pm 5.9$ \\ \hline
\end{tabular}
\end{center}
\label{tab:data}
\end{table}
The uncertainties are only statistical.

	We discuss the sources of systematic uncertainties on our measurements
of the width of the $D^{\ast +}$ in
the order of their size.  The most important contribution is the variation
of the results as a function of the kinematic parameters of
the $D^{\ast +}$ decay.
The next most important contribution
comes from any mismodeling of $\sigma_Q$'s
dependence on the kinematic parameters.
We take into account correlations among the
less well measured parameters of the fit, such as $f_{mis}$ and
$\sigma_{mis}$,
by fixing each parameter at $\pm 1\sigma$ from their central fit values,
repeating the fit, and adding in quadrature the variation in
the width of the $D^{\ast +}$ and $Q_0$ from their central values.
We have studied in the simulation the sources of
mismeasurement that give rise
to smearing on the width of the $D^{\ast +}$ by replacing the
measured values with the generated values for various kinematic
parameters of the decay products.  We have then compared
these uncertainties with analytic expressions for the uncertainties.
The only source of smearing that we cannot account for in
this way is a small distortion of the kinematics of the event
caused by the algorithm used to reconstruct the $D^0$ origin point
described above.
We have also checked that our simulation accurately models the line shape
of other narrow resonances visible in our data.  Notably the 
decay $\Lambda^0 \to p\pi^-$, has a $Q$ only seven times that of
$D^{\ast+} \to D^0 \pi^+_{\rm slow}$.  In the $\Lambda^0$ decay we
select the $\pi^-$ to have a momentum in the range of those in
the $D^{\ast+}$ decay, and the visible $\Lambda^0$ widths agree to a
few percent between data and simulation.
We consider uncertainties from the background shape by allowing the
coefficients of the background polynomial to float. 
Minor sources of uncertainty are from the width offsets derived
from our simulation and given in Table~\ref{tab:data}, and
our digitized data storage format.

	An extra and dominant source of uncertainty on $Q_0$ is the energy
scale of our measurements.
We evaluate this uncertainty by studying
$K_s \to \pi^+\pi^-$ decays in our data.
In order to bring the $K_s$ mass central value in agreement with the nominal
one, we make small relative momentum corrections, less than
0.3\%, for tracks with momenta between 100 and 500 MeV/c.
These corrections only 
affect the slow pion.  Applying these corrections to the momentum
of the slow pion in our data we find a shift in the fit
value of $Q_0$, $-4$ keV for all the samples, and
a negligible change in the width.
We evaluate uncertainties
in the energy scale by varying an overall momentum scale to
change the $K_s \to \pi^+\pi^-$ mass
by $\pm30$ keV, the uncertainty on that mass \cite{PDG},
and applying the statistical errors we obtain
from the calculations of the momentum corrections discussed above.

	Table~\ref{tab:systematic} summarizes the systematic
\begin{table}
\caption{Systematic uncertainties on the width of the $D^{\ast +}$ and $Q_0$}
\begin{center}
\begin{tabular}{|c|c|c|c|c|c|c|} \hline
                         & \multicolumn{6}{c|}{Uncertainties in keV} \\
                         & \multicolumn{6}{c|}{Sample} \\ \hline
                         & \multicolumn{2}{c|}{Nominal}
                         & \multicolumn{2}{c|}{Tracking}
                         & \multicolumn{2}{c|}{Kinematic} \\ \hline

Source                    & $\delta \Gamma(D^{\ast +})$ & $\delta Q_0$
                           & $\delta \Gamma(D^{\ast +})$ & $\delta Q_0$
                           & $\delta \Gamma(D^{\ast +})$ & $\delta Q_0$ \\ \hline
Dependence on Kinematics  & 16 &  8   & 16 &  8   & 16 &  8 \\
Mismodeling of $\sigma_Q$ & 11 & $<1$ &  9 &  4   &  7 & $<1$ \\
Fit Correlations          &  8 &  3   &  9 &  4   &  9 &  5 \\
Vertex Reconstruction     &  4 &  2   &  4 &  2   &  4 &  2 \\
Background Shape          &  4 & $<1$ &  2 & $<1$ &  2 & $<1$ \\
Offset Correction         &  2 &  3   &  6 & 10   &  3 &  5 \\
Data Digitization         &  1 &  1   &  1 &  1   &  1 &  1 \\ 
Energy Scale              &  1 &  8   &  1 &  8   &  1 &  8 \\ \hline
Quadratic Sum             & 22 & 12   & 22 & 16   & 20 & 14 \\ \hline
\end{tabular}
\end{center}
\label{tab:systematic}
\end{table}
uncertainties on the width of the $D^{\ast +}$ and $Q_0$.

	In summary we have measured the width of the
$D^{\ast +}$ by studying the
distribution of the energy release in $D^{\ast +} \to D^0 \pi^+$ followed
by $D^0 \to K^- \pi^+$ decay.
With our estimate of the systematic uncertainties for each of the
three samples being essentially the same we chose to report the result
for the sample with the smallest statistical uncertainty, the minimally
selected sample, and obtain
\begin{equation}
\Gamma(D^{\ast +}) = 96 \pm 4 \pm 22\ {\rm keV},
\label{eq:result}
\end{equation}
where the first uncertainty is statistical and the second is systematic.
This is the first measurement of the width of
the $D^{\ast +}$, and
it corresponds to a strong coupling\cite{pred}
\begin{equation}
g = 0.59 \pm 0.01 \pm 0.07.
\end{equation}
This is consistent with theoretical predictions based on HQET and
relativistic quark models, but higher than predictions based on QCD
sum rules.  
We also measure the mean value for the energy release in
$D^{\ast +} \to D^0 \pi^+$ decay
\begin{equation}
Q_0 = 5842 \pm 2 \pm 12\ {\rm keV},
\end{equation}
where the first error is statistical and second is systematic.
Combining this with the mass of the charged pion, 139.570 MeV with
an uncertainty less than 1 keV \cite{PDG}, we calculate
\begin{equation}
m_{D^\ast(2010)^+} - m_{D^0} = 145.412 \pm 0.002 \pm 0.012\ {\mathrm MeV}.
\end{equation}
This agrees with the value from the Particle Data Group,
$145.436 \pm 0.016$ MeV,
from a global fit of all flavors of $D^\ast$--$D$ mass differences.

\section*{Acknowledgments}

We thank D.~Becirevic,
I.~I.~Bigi, G.~Burdman, A.~Khodjamirian, P.~Singer, and A.~L.~Yaouanc
for valuable discussions.
We gratefully acknowledge the effort of the CESR staff in providing us with
excellent luminosity and running conditions.
This work was supported by 
the National Science Foundation,
the U.S. Department of Energy,
the Research Corporation,
the Natural Sciences and Engineering Research Council of Canada
and the Texas Advanced Research Program.

\end{document}